\documentclass{aa}  
\usepackage{txfonts}
\usepackage{graphicx}
\usepackage{natbib}

\begin{document}

   \title{Long-baseline optical intensity interferometry}

   \subtitle{Laboratory demonstration of diffraction-limited imaging}

   \author{Dainis Dravins
          \inst{1},
               Tiphaine Lagadec,
           \inst{1,2}
          \and
          Paul D.\ Nu{\~n}ez
          \inst{3,4,5} \fnmsep 
		}
%
%
   \institute{Lund Observatory, Box 43, SE-22100 Lund, Sweden\\
              \email{dainis@astro.lu.se}
\and
     Present address: ESTEC, European Space Research and Technology Centre, Keplerlaan 1, NL-2200 AG  Noordwijk, \\ The Netherlands\\
              \email{lagadec.tiphaine@gmail.com}
\and
             Laboratoire Lagrange, Observatoire de la C\^ote d'Azur, BP 4229, FR-06304 Nice, France
 \and
		Coll\`ege de France, 11 Place Marcelin Berthelot, FR-75005 Paris, France
\and 
		Present address: Jet Propulsion Laboratory, California Institute of Technology, 4800 Oak Grove Drive, Pasadena, \\ CA 91109-8099, U.S.A.\\
		\email{paul.nunez@jpl.nasa.gov}
             }

   \titlerunning{Long-baseline optical intensity interferometry}
   \authorrunning{D.\ Dravins et al.}  

   \date{Received April 16, 2015; accepted May 22, 2015}

 
  \abstract
  {A long-held vision has been to realize diffraction-limited optical aperture synthesis over kilometer baselines. This will enable imaging of stellar surfaces and their environments, and reveal interacting gas flows in binary systems. An opportunity is now opening up with the large telescope arrays primarily erected for measuring Cherenkov light in air induced by gamma rays. With suitable software, such telescopes could be electronically connected and also used for intensity interferometry.  Second-order spatial coherence of light is obtained by cross correlating intensity fluctuations measured in different pairs of telescopes.  With no optical links between them, the error budget is set by the electronic time resolution of a few nanoseconds. Corresponding light-travel distances are approximately one meter, making the method practically immune to atmospheric turbulence or optical imperfections, permitting both very long baselines and observing at short optical wavelengths.}
   {Previous theoretical modeling has shown that full images should be possible to retrieve from observations with such telescope arrays.  This project aims at verifying diffraction-limited imaging experimentally with groups of detached and independent optical telescopes.}
   {In a large optics laboratory, artificial stars (single and double, round and elliptic) were observed by an array of small telescopes.  Using high-speed photon-counting solid-state detectors and real-time electronics, intensity fluctuations were cross-correlated over up to 180 baselines between pairs of telescopes, producing coherence maps across the interferometric Fourier-transform plane. }
   {These interferometric measurements were used to extract parameters about the simulated stars, and to reconstruct their two-dimensional images.  As far as we are aware, these are the first diffraction-limited images obtained from an optical array only linked by electronic software, with no optical connections between the telescopes.} 
   {These experiments serve to verify the concepts for long-baseline aperture synthesis in the optical (somewhat analogous to radio interferometry) and to optimize the instrumentation and observing procedures for future observations with large arrays of Cherenkov telescopes, aiming at order-of-magnitude improvements of the angular resolution in optical astronomy.}

   \keywords{Intensity interferometry -- Hanbury Brown - Twiss -- Aperture synthesis -- Second-order coherence -- Stellar surface imaging -- Exoplanet imaging -- Cherenkov telescopes }

   \maketitle

\section{Introduction}

Highest angular resolution currently realized in optical astronomy is offered by amplitude (phase-) interferometers that combine light from telescopes separated by baselines of up to a few hundred meters. Tantalizing results show how stellar disks start to become resolved, revealing stars as a diversity of individual objects, although so far feasible only for a few of the largest ones. Several concepts have been proposed to extend such facilities to scales of a kilometer or more, but their realization remains challenging either on the ground or in space. Limiting parameters include the requirement of optical and atmospheric stability to a fraction of an optical wavelength, and also the need to cover many interferometric baselines, given that optical light cannot be copied with retained phase, but has to be split up (and diluted) by beamsplitters to achieve interference among multiple telescope pairs.  Overviews of those techniques are given by \citet{Millour_2008} and \citet{Quirrenbach_2009} and in more detail by \citet{Monnier_2003} and \citet{Monnier_Allen_2013}, while recent results are highlighted in \citet{Berger_etal_2012} and \citet{vanBelle_2012}.

Bright stars typically subtend diameters of only a few milliarcseconds (mas), requiring interferometry over many hundreds of meters or some kilometer to enable surface imaging. Telescope complexes of such an extent are now being constructed on the ground as arrays of air Cherenkov telescopes. These are erected to measure brief flashes of visual Cherenkov light in air induced by energetic gamma rays but their optomechanical performance is fully adequate for intensity interferometry also, a technique already pioneered long ago \citep{Hanbury_Brown_1974}. Because it is essentially insensitive to atmospheric turbulence, this method permits both very long baselines and observing at short optical wavelengths; only the lack of suitably large and well-distributed telescopes seem to have caused the method not to be pursued recently in astronomy.

Currently, the largest such project is CTA, the Cherenkov Telescope Array, foreseen to have around 100 telescopes spread over a few square kilometers \citep{Acharya_etal_2013, CTA_2015}. Among its envisioned uses is also that of an intensity interferometer, correlating intensity fluctuations between numerous telescope pairs across different baselines while applying timeshifts in software to track sources across the sky \citep{Dravins_etal_2012, Dravins_etal_2013}. Since the signals are copied electronically (with no optical connection between telescopes), there is no loss of signal when forming additional baselines between any pairs of the numerous telescopes.

Theoretical and numerical simulations of such observations have been made for several types of objects \citep{Nunez_etal_2012a, Nunez_etal_2012b, Rou_etal_2013, Dolne_etal_2014}. Here, we report on their experimental verification.  A laboratory setup was prepared with artificial optical sources (’stars’), observed by an array of small telescopes equipped with nanosecond-resolving photon-counting solid-state detectors. The photon streams (reaching MHz levels, as for realistic telescope operations) were fed into digital firmware correlators in real time (with nanosecond resolution, as required for actual telescope operations), computing cross correlations between the `random' intensity fluctuations simultaneously measured in many different pairs of telescopes. The degree of mutual correlation for any given baseline provides a measure of the second-order spatial coherence of the source at the corresponding spatial frequency (and thus the square of the ordinary first-order coherence and the square of the Fourier transform component of the source’s brightness distribution). Numerous telescope pairs of different baseline lengths and orientations fill in the interferometric (u,v)-plane, and the measured data points build up a two-dimensional map of the second-order spatial coherence of the source, from which its image can be extracted. The experience from such laboratory experiments will serve to verify instrumental concepts and serve as input for specifying the procedures for future full-scale observations with large telescope arrays.

\section{Principles of intensity interferometry} 

The technique of intensity interferometry was developed for the original purpose of measuring stellar sizes, and a dedicated instrument was built at Narrabri, Australia \citep{Hanbury_Brown_1974}.  What is observed is the second-order coherence of light (i.e., that of intensity $I$, not of amplitude or phase) by measuring temporal correlations of arrival times between photons recorded in different telescopes.  A star is observed with two separate telescopes, recording the (quantum-optically) random and very rapid intrinsic fluctuations in starlight (fully resolved on timescales equal to the temporal coherence of light, perhaps $10^{-13}$ s).  With the telescopes close together, the fluctuations in both telescopes are correlated in time, but lose correlation for larger separations. 

The quantity measured is
\begin{equation}
 \langle I_1(t)\cdot I_2 (t) \rangle = \langle I_1(t) \rangle \langle I_2(t) \rangle (1 + |\gamma_{12}|^2),
\end{equation}

\noindent  where  $\langle \rangle$ denotes temporal averaging and $\gamma_{12}$ is the mutual coherence function of light between locations 1 and 2, the quantity commonly measured in phase/amplitude interferometers (a relation that holds for each polarization). Compared to randomly fluctuating intensities, the correlation between intensities $I_1$ and $I_2$ is `enhanced' by the coherence parameter and an intensity interferometer thus measures $|\gamma_{12}|^2$ with a certain electronic time resolution. This relation holds for ordinary thermal (chaotic, maximum-entropy, Gaussian) light, where the wave undergoes random phase jumps over timescales of its coherence time but not necessarily for light with different photon statistics (e.g., an ideal laser emits coherent light of constant intensity, without any phase jumps, and thus would not generate any reasonable signal in an intensity interferometer).  

The great observational advantage of intensity interferometry is that it is practically insensitive to either atmospheric turbulence or to telescope optical imperfections, enabling very long baselines and observations at short optical wavelengths, even through large airmasses far away from the zenith. Since telescopes are only connected electronically, error budgets and required precisions relate to electronic timescales of nanoseconds, and light-travel distances of tens of centimeters or meters rather than small fractions of an optical wavelength.  For more detailed discussions of the principal workings of intensity interferometry, see monographs by \citet{Labeyrie_etal_2006}, \citet{Saha_2011}, or \citet{Shih_2011}, for example.

There is a price to be paid, however, for this freedom from atmospheric influences. Realistic time resolutions are much longer than typical optical coherence times for broad-band light, and any measured intensity-fluctuation signal is averaged over very many coherence times. Therefore, very good photon statistics are necessary for a reliable determination of the smeared-out signal, requiring large photon fluxes (thus large telescopes); already the 6.5 m flux collectors in the original intensity interferometer at Narrabri were larger than any other optical telescope at that time.  


The technical specifications for air Cherenkov telescopes such as the forthcoming Cherenkov Telescope Array \citep{Acharya_etal_2013, CTA_2015}, are remarkably similar to the requirements for intensity interferometry. In the original Narrabri instrument, the telescopes were moved on tracks during observation to maintain their projected baseline; however electronic time delays can now be used instead to compensate for different arrival times of a wavefront to the different telescopes, removing the need for having them mechanically mobile.

With perhaps 100 telescopes, the CTA will provide an unprecedented light-collecting area of some 10,000 m$^2$ distributed over a few square kilometers. Of course, it will mainly be devoted to its primary task of observing Cherenkov light in air; however several other applications have been envisioned, to be preferably carried out during nights with bright moonlight which -- owing to the faintness of the Cherenkov light flashes -- might preclude their efficient observation. Besides intensity interferometry \citep{Dravins_etal_2013}, additional uses that have been suggested include searches for rapid astrophysical events \citep{Deil_etal_2009, Hinton_etal_2006, Lacki_2011}, observations of stellar occultations by distant Kuiper-belt objects \citep{Lacki_2014}, or as a terrestrial ground station for optical communication with distant spacecraft \citep{Carrasco-Casado_etal_2013}. For practical operations, one proposed concept is to place auxiliary detectors on the outside of the Cherenkov camera cover lid, an approach already realized on some telescopes \citep{Deil_etal_2009, Hinton_etal_2006}, and one that should not interfere with regular Cherenkov camera operations.

This potential of using Cherenkov telescope arrays for optical intensity interferometry has indeed been noticed by several \citep{LeBohec_Holder_2006, LeBohec_etal_2008}. If, for example, a baseline of 2 km could be utilized with CTA at $\lambda$~350 nm, resolutions would approach 30 $\mu$as, which is an unparalleled  spatial resolution in optical astronomy \citep{Dravins_etal_2012, Dravins_etal_2013}. 

\section{Verifying intensity interferometry}

Several studies have considered the observational challenges, such as the overall sensitivity and limiting stellar magnitudes, and those issues now appear to be consistently and generally understood \citep{Dravins_etal_2012, Dravins_etal_2013, LeBohec_Holder_2006, Trippe_etal_2014}.  For example, current types of electronics with current Cherenkov telescope designs should, during one night of observing in one single wavelength band, permit measurements of hotter stars down to a visual magnitude of about m${_V}$=8, giving access to many thousands of sources. 

Comprehensive Monte Carlo simulations were done by \citet{Rou_etal_2013}. They numerically simulated effects of various noise sources, of background light from the night sky, effects of large collecting areas (i.e., non-negligible in comparison to the coherence pattern), among others. For example, the effect of large-size telescopes is that the signal will correspond to a convolution of the degree of coherence $|\gamma|^2$ with each of the telescope light-collection area shapes. However, in real observing, there will also arise various experimental issues, not all of which are still understood well. Some were examined already long ago during the Narrabri interferometer operations, while others are specific to modern detectors and electronics (dead-time or saturation effects in detectors and digital data handling). 

\subsection{Experimental test observations}

Full-scale experiments in connecting major Cherenkov telescopes for intensity interferometry were made by \citet{Dravins_LeBohec_2008}, using pairs of 12-m telescopes of the VERITAS array on Mt.Hopkins, Arizona, at baselines between 34 and 109 m. In each telescope, starlight was recorded by one photomultiplier in the center of the regular Cherenkov cameras, its signal digitized, and sent via optical cables to the control building, where they were fed into a real-time digital correlator at continuous photon-count rates up to some 30 MHz. Although no astrophysical correlations were recorded, these experiments verified that there seemed to be no particular operational problems in using large Cherenkov telescopes for intensity interferometry.

Significant efforts have been made at The University of Utah in developing experimental test facilities for intensity interferometry, including 3-m Cherenkov telescopes at its `StarBase' facility and laboratory setups \citep{LeBohec_etal_2010}.  For example, an analog correlator based upon a Field Programmable Gate Array (FPGA) was used for laboratory tests \citep{Nunez_2012} with intensity fluctuations in starlight mimicked by a pseudo-thermal light source, which is produced by shining a laser through a rotating ground-glass plate, producing a time-variable speckle pattern \citep{Martienssen_Spiller_1964}.  

Also \citet{Pellizzari_etal_2012} used a rotating ground-glass plate to create speckle patterns in the observation plane with two laboratory telescopes. Cross correlating the light intensity between these provides a measure of the speckle extent, hence the source size. These experiments have provided insights into detector properties and signal handling although this type of light sources do not have the small angular extents that would be appropriate for astronomical imaging. 

\subsection{Photon-counting detectors}

Air Cherenkov telescopes have traditionally been equipped with photomultiplier (PMT) detectors, a natural choice since the image point-spread-function in the focal plane is typically some cm in extent, so well matched to PMT sizes. However, their quantum efficiency is not perfect, and their electric requirements and consequences from overexposure may cause some concern. Recently, higher-efficiency and less fragile silicon avalanche photodiode arrays have been incorporated into Cherenkov cameras \citep{Anderhub_etal_2013, Catalano_etal_2013, Sottile_etal_2013} and it can be envisioned that such detectors will also be used in future telescopes, at least those with an optical two-mirror design that produce a fairly compact focal plane.

Single-photon-counting silicon avalanche photo-diodes (SPADs) have a potential for quantum efficiency that approaches unity (and extends into the infrared), while counting individual photons at nanosecond resolution. They are operated in `Geiger mode' where each detected photon triggers an electronic avalanche as a signature of photon detection. Following a detection, a certain deadtime occurs, during which further detections are not possible. Typical deadtimes can be tens of nanoseconds, which permit count rates up to 10 MHz or somewhat higher. 

In addition to dark counts, detectors also show some level of afterpulsing. There remains a possibility that an avalanche electron is occasionally caught in the potential well around some semiconductor impurity site. If that trapped electron is released after a time longer than the deadtime, it may trigger a new avalanche, which is correlated with the real photon event. Since, in intensity interferometry, one searches for correlated signals, such afterpulsing may appear as a false correlation. Of course, the afterpulsing is a property inside each single detector and will not (except as a small higher-order effect) be correlated between two separate ones.

Another somewhat peculiar property of SPADs is the emission of light from the detector surface (`diode afterglow').  Immediately following the photon-detecting avalanche, a shower of photons is emitted from the detector as it recovers. For high-speed applications in large telescopes, some such secondary light could find its way back onto the detector, causing optical `ringing'. This emission can also induce crosstalk between adjacent pixels in a detector array (`optical crosstalk'). In the electronic design of SPAD arrays, efforts have been made to remedy such effects by, for instance, making trenches in the silicon to avoid direct exposure between adjacent pixels. 

These examples illustrate that an adequate understanding of detector properties is desirable in both the selection and the use of actual detectors. Since solid-state detectors appear to be a promising choice in terms of sensitivity and handling, those were chosen for the present experiments. One practical concern is that the physical size of single-pixel detectors is tiny, at only a fraction of a mm, so incompatible with the large image scales in Cherenkov telescopes, where larger array detectors have to be used. Large solid-state photomultiplier arrays are now available from several manufacturers but already the experience from smaller-scale ones should be relevant for future full-scale operations. 

Single-photon-counting avalanche silicon photodiode detector modules (SPADs) from various manufacturers were evaluated in laboratory tests. Following those, a dozen modules from MPD, Micro Photon Devices, with sensitive areas of 100~$\mu$m diameter, Peltier-cooled, and with dark-count rates below 500~Hz were acquired to equip a corresponding number of small laboratory telescopes.

\subsection{Real-time photon correlation}

An essential element of an intensity interferometer is the correlator, which provides the temporally averaged product of the intensity fluctuations  $\langle I_1\cdot I_2 \rangle$ from two telescopes, normalized by the product of average intensities in each of them,  $\langle I_1\rangle \cdot \langle I_2 \rangle$.  The original interferometer at Narrabri used an analog correlator to multiply the photocurrents from its photomultipliers. Analog correlators \citep{LeBohec_etal_2008, Nunez_2012, Pellizzari_etal_2012} have attractive properties in often being able to handle wide electrical bandwidths. However, they may also be sensitive to electromagnetic interference from all sorts of electrical devices omnipresent also in observatory environments \citep{Pellizzari_etal_2012}. For applications in intensity interferometry, these may be critical since the measured signal is expected to be tiny, and has to be segregated from a very much greater background. Of course, isolation from interference is feasible with sufficient shielding, and having optical cables instead of metallic ones may alleviate the problems, but there are still issues at their interfaces. Because of these considerations, the signal handling was chosen here to only be digital. This may introduce limitations in how high count rates that can be handled per channel but -- since the S/N in principle is independent of the optical bandpass (see below) -- stellar fluxes may be adjusted to manageable photon-count levels using suitably narrow-band optical filters.

Photon correlators are commercially available for primary applications in light scattering from laboratory specimens \citep{Becker_2005}.  Based upon experiences from a series of hardware correlators from different manufacturers, current units were custom-made (by {\it Correlator.com}) for applications in intensity interferometry.  These feature sampling frequencies up to 700 MHz, are able to handle continuous photon-count rates of more than 100 MHz per channel without any deadtimes, and have on-line data transfer to a host computer. Their output contains the cross correlation function between pairs of telescopes (as well as autocorrelation functions for each of them), made up of about a thousand points. For small delays, the sampling is made with the shortest timesteps of just over 1 ns, increasing in a geometric progression to high values to reveal the full function up to long delays of even seconds or more. It is believed that their performance is adequate for full-scale intensity interferometry experiments. In particular, it should be stressed that the correlators required here are very much more modest than those used in radio phase interferometer facilities (of supercomputer class), which decipher not only spatial but also spectral information, and with many bits of resolution.

Still, such correlators are not without problems. In the ideal case of a spatially coherent source measured with perfect temporal resolution, the average of the instantaneously squared intensity equals twice the square of its average (Eq.1).  To obtain such a value in practice requires measurements with a time resolution shorter than the optical coherence time.  While this can be realized for special laboratory light sources (as discussed below), it is not realistic for astronomical ones (except perhaps some hypothetical natural lasers), where the optical passband is defined by some wavelength filter.  Realistic bandpasses of 1 nm or 10 nm correspond to optical frequency bandpasses on the order of 1 or 10 PHz and coherence times around one picosecond or less.  Since realistic electronic resolutions are close to a nanosecond, the intrinsic intensity fluctuations are averaged over many optical coherence times, and the measured squared intensity deviates only slightly from the square of its average, rather than by the full factor of two.   With random fluctuations decreasing as the square root of the number of samples, this smearing may be a factor of 100 or more, requiring both large collecting areas to reach adequate photon statistics, and analysis of the correlation signal to many decimal digits.

Firmware correlators produce correlation functions in real time, processing very large amounts of photon-count data and eliminating the need for their further handling and storage: for example, ten input channels, each running at 50 MHz during one eight-hour observing night, process more than 10 terabytes of photon-count data). A disadvantage is that, if something needs to be checked afterward, the original data are no longer available, and alternative signal processing cannot be applied.

An alternative approach is to time-tag each photon count and store all data (at least for moderate count rates), and then later perform the correlation analyses off-line. The data streams from multiple telescopes can then be cross-correlated in software, possibly applying noise filters and also computing other spatio-temporal parameters such as higher-order correlations between three or more telescopes, which could contain additional information. Such a capability was built into the {\it AquEYE} and {\it IquEYE} instruments developed at the University of Padova for very high time-resolution astrophysics, also with a view toward intensity interferometry, using similar SPAD detectors to those here \citep{Capraro_etal_2010, Naletto_etal_2009, Naletto_etal_2013}.

\subsection{Image reconstruction from second-order coherence}

While intensity interferometry possesses the advantage of not being sensitive to phase errors in the optical light path, ordinary two-telescope correlations also do not permit such phases of the complex coherence to be measured. These correlations provide the absolute magnitudes of the respective Fourier transform components of the source image, while the phases are not directly obtained. Such quantitites can be used well by themselves to fit model parameters such as stellar diameters, stellar limb darkening, binary separations, and circumstellar disk thicknesses, but actual images cannot be directly computed through a simple inverse Fourier transform.

While a two-component interferometer (such as the classical one at Narrabri) offers only very limited coverage of the Fourier (u,v)-plane, a multicomponent system provides numerous baselines and an extensive coverage of the interferometric plane. Already intuitively, it is clear that the information contained there must place stringent constraints on the source image. For instance, viewing the familiar Airy diffraction pattern (cf.\ Figure 1 left), one immediately recognizes it as originating in a circular aperture, although only intensities are seen. However, it is also obvious that a reasonably complete coverage of the diffraction image is required to convincingly identify a circular aperture as the source. 

Various techniques (most unrelated to astronomy) have been developed for recovering the phase of a complex function when only its magnitude is known. Methods specifically for intensity interferometry were worked out by Holmes et al.\ (2004; 2010; 2013) for one and two dimensions, respectively. Once a sufficient coverage of the Fourier plane is available, phase recovery and imaging indeed become possible. \citet{Nunez_etal_2012a, Nunez_etal_2012b} applied this phase recovery to reconstruct images from simulated intensity interferometry observations, demonstrating that also rather complex images can be reconstructed. (However, a limitation that still remains is the non-uniqueness between the image and its mirrored reflection.)

\section{Telescope array in the laboratory}

The project aim was to realize an end-to-end simulation of an intensity interferometer observing star-like sources with a large array of many telescopes, equipped with electronics of the type foreseen for full-scale operations. An interferometer was thus set up in a very large optics laboratory, comprising an artificial ‘star’ viewed by a linear array of typically five stationary telescopes of different separations. A two-dimensional telescope layout with many tens of telescopes could be simulated by successively rotating the position angle of the ‘star’ relative to the plane of the telescopes. With high-speed photon-counting detectors and real-time digital cross correlation between many tens or even hundreds of telescope pairs, operations resemble foreseen future observations with large Cherenkov telescope arrays. Still, such a laboratory setup demands several considerations.

   \begin{figure}
   \centering
   \includegraphics[width=\hsize]{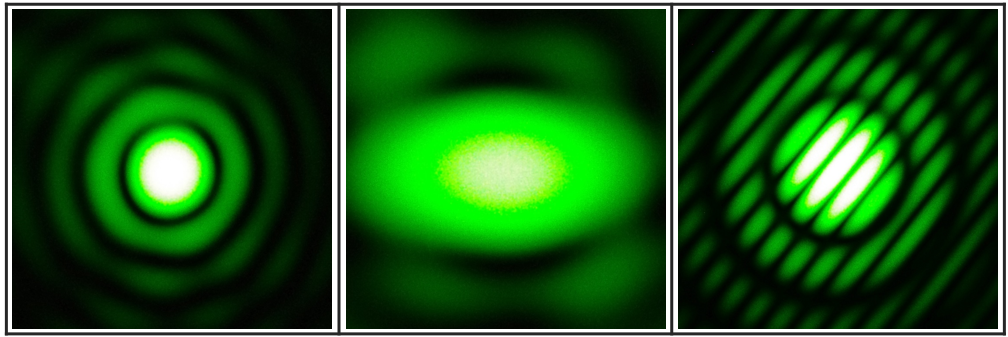}
      \caption{Diffraction patterns in $\lambda$~532 nm laser light show the [squared] Fourier transforms of some of the artificial `stars'. Left to right: Circular single star of 150 $\mu$m nominal diameter; elliptical small single star; symmetric binary star with each component 100 $\mu$m. The side of each image corresponds to $\sim$~70 cm in the telescope plane and corresponding baselines are required to retrieve these patterns. }
         \label{FigDiffraction}
    \end{figure}

\subsection{Small sources and long distances}

A first issue concerns the artificial `stars'. These should have somewhat realistic angular sizes and also possess some nontrivial properties whether single (round or elliptical), or double (with equal or unequal components). They were prepared as small apertures, drilled as physical holes in metal (using miniature drills otherwise intended for mechanical watches). This sets some minimum practical size of aperture openings of $\sim$100$~\mu$m. Such an opening observed at 20 m distance subtends an angle of 5$\cdot10^{-6}$ rad (1 arcsec). With the Airy disk diffraction radius given by $\Theta \approx 1.22 \lambda/D$ rad, the telescope aperture D required to resolve it comes out to $\sim$10 cm, dictating a rather compact setup of telescopes unless extremely large laboratory spaces are available.  Our interferometer occupied most of the wall-to-wall extent of the large optics laboratory at Lund Observatory, with a source-to-telescope distance of $\sim$23 m.  This choice followed an evaluation of alternative setups simulated with optical design software ({\it{ZEMAX}}), where we found that the requirements of source surface brightness and small optical aberrations to match a small detector size were too awkward to satisfy with variants involving, for example, multiple-reflection passes, optical fibers, or large-size lenses.

Figure 1 shows photographs of the diffraction patterns of some of the artificial stars, here illuminated by coherent laser $\lambda$~532 nm light, projected onto a large screen. These patterns – Fourier transforms of the aperture openings – serve both to obtain precisely the dimensions and shapes of the `stars' (quantities otherwise difficult to directly measure on the tiny openings) and to illustrate the spatial coherence patterns that can eventually be retrieved from interferometry. 

\subsection{Dependence on source temperature}

The signal-to-noise ratio (S/N) in measuring the second-order coherence for one pair of telescopes equals 
\begin{equation}
(S/N)_{RMS}= A \cdot \alpha \cdot \eta \cdot |\gamma_{12}(\mathbf{r})|^2 \cdot \Delta f^{1/2} \cdot (T/2)^{1/2}
\end{equation}

\noindent e.g., \citet{Hanbury_Brown_1974}. Here $A$ is the geometric mean of the areas of the two telescopes; $\alpha$ is the quantum efficiency of the optics plus detector system; $\eta$ is the flux of the source in photons per unit optical bandwidth, per unit area, and per unit time; $|\gamma_{12}(\mathbf{r})|^2$ is the second-order coherence of the source for the baseline vector $\mathbf{r}$, with $\gamma_{12}(\mathbf{r})$ being the mutual degree of coherence.  And $\Delta f$ is the electronic bandwidth of the detector plus signal-handling system, and $T$ the integration time.

Most of these parameters are easily understood since they directly depend on the instrumentation, but $\eta$ depends on the source itself, because it is a function of its brightness temperature.  Thus, for a given number of photons detected per unit area and unit time, the S/N is better for sources where those photons are squeezed into a narrower optical band. This implies a valuable property for intensity interferometry in that the S/N is independent of the width of the optical passband $\Delta\lambda$, as shown by Eq. (2).   While a smaller flux over a narrower passband worsens the photometric precision due to photon statistics (assuming a flat source spectrum), that is counterbalanced by a longer optical coherence time, and a diminished smearing of the physical signal during the electronic time resolution.

This property has already been exploited in the original Narrabri interferometer \citep{Hanbury_Brown_etal_1970} to identify the extended emission-line volume from the stellar wind around the Wolf–Rayet star $\gamma^2$~Vel.  They compared observations in the $\lambda$~465 nm \ion{C}{iii-iv} emission feature emanating from the wind, with those in the photospheric continuum around $\lambda$~443 nm, finding the extent of the emission-line region several times greater than the stellar disk, and concluding that the stellar wind fills the entire Roche equipotential lobe around this binary star.  Their continuum spectral region (defined by a 10 nm wide color filter) was selected to be free from prominent spectral features, while the ionized carbon emission was measured over a narrower band of 2.5 nm. The same effect could also be exploited to increase the S/N by observing the same source simultaneously in multiple spectral channels, a concept foreseen for the once-proposed successor to the original Narrabri interferometer \citep{Hanbury_Brown_1991}. 

That the limiting S/N is set by the source temperature rather than by its brightness or telescope size, may appear somewhat counter-intuitive (and has been the cause of some misunderstandings) but can be interpreted in both classical and quantum terms. Once the observational equipment is in place, one may try to improve the S/N. One could try increasing the photon flux by going to a broader wavelength interval in white light. However, the S/N will not change: realistic electronic resolutions ($\sim$ns) are always very much slower than the temporal coherence time of broad-band light (perhaps $10^{-14}$s). While broadening the spectral passband does increase the photon count rate, it also decreases the temporal coherence by the same factor. The intensity fluctuations have their full amplitude over one coherence time, and now get averaged over many more, canceling the effects of decreased photon noise. Then one could try with larger telescopes. However, once telescope sizes begin to approach the structures in the spatial coherence pattern, although increasing telescope areas gives more photons, they also average over more spatial coherence structures by the same factor, again not improving the S/N. Alternatively, one might be tempted to observe brighter sources. However, for any given source temperature, brighter sources will be larger in angular extent and have smaller structures in the Fourier domain, so one will again average over additional spatial coherence areas, without improving the S/N. Indeed, even pointing the telescope at the Sun will not help since the only way to get better S/N is to find sources with a higher surface brightness. Below some effective source temperature, no sensible measurements can be achieved, no matter how bright the source, or how large the telescopes.  A diagram illustrating this steep temperature dependence is in \citet{Dravins_Lagadec_2014}; their Figure 3. 

\subsection{High-temperature laboratory light sources}

That intensity interferometry is sensitive to sources of high brightness temperature but limited in observing cool ones, of course, is equally valid for any laboratory setup, as for stars in the sky. The source must be small enough to produce an extended Fourier pattern that can be sampled by the interferometer, while also being bright enough to produce acceptable photon count rates. However, while there are many stars in the sky with T$_{\rm{eff}}$ = 10,000~K or more, to produce a correspondingly brilliant laboratory source is much more challenging. Furthermore, the precise relation between second- and first-order coherence, $\varg^{(2)} = 1 + |\varg^{(1)}|^2$, assumes that the light is chaotic, with a Gaussian amplitude distribution \citep{Shih_2011, Bachor_Ralph_2004, Loudon_2000}, i.e., the light waves undergo random phase shifts so that intensity fluctuations result, which then bear a simple relation to the ordinary first-order coherence. While this must be closely satisfied for any thermal source, it is not the case for lasers that, ideally, never undergo any intensity fluctuations anywhere ($\varg^{(2)}$ = 1). The spatial extent of a laser source therefore cannot be measured by intensity interferometry, and a laser is not an option for enhancing the brightness of an artificial star.

In initial attempts to achieve a high surface brightness for an artificial ‘star’, the very small emission volume of a high-pressure Hg arc lamp was focused onto a pinhole serving as the ‘star’, and a narrow-band optical filter singled out its brightest emission line ($\lambda$~546 nm). Although this represents about the highest blackbody brightness temperature (some 5,500 K) that can be obtained with somewhat ordinary laboratory equipment for a non-laser source, the photon-count rates still turned out to be too low for conveniently short integration times. Other tests were made with several lamps of various atomic species, selecting their brightest and narrowest emission lines, but the signals still remained marginal. Finally, quasi-monochromatic chaotic light was produced by scattering monochromatic laser light from microscopic particles, suspended in a water-filled cuvette. Such particles undergo random thermal (Brownian) motion due to collisions with the water molecules that surround them, producing a slightly Doppler-broadened spectral line that is extremely narrow ($\sim$~MHz). Thus, a light source with an effective brightness temperature estimated as T$_{\rm{eff}}$ $\sim$100~000 K was produced, permitting conveniently short integration times on the order of a minute. Such scattered laser light is used for various laboratory photon-correlation measurements of time variability \citep{Becker_2005}, but we are not aware of any previous such experiment with a spatial intensity interferometer.

   \begin{figure}
   \centering
   \includegraphics[width=\hsize]{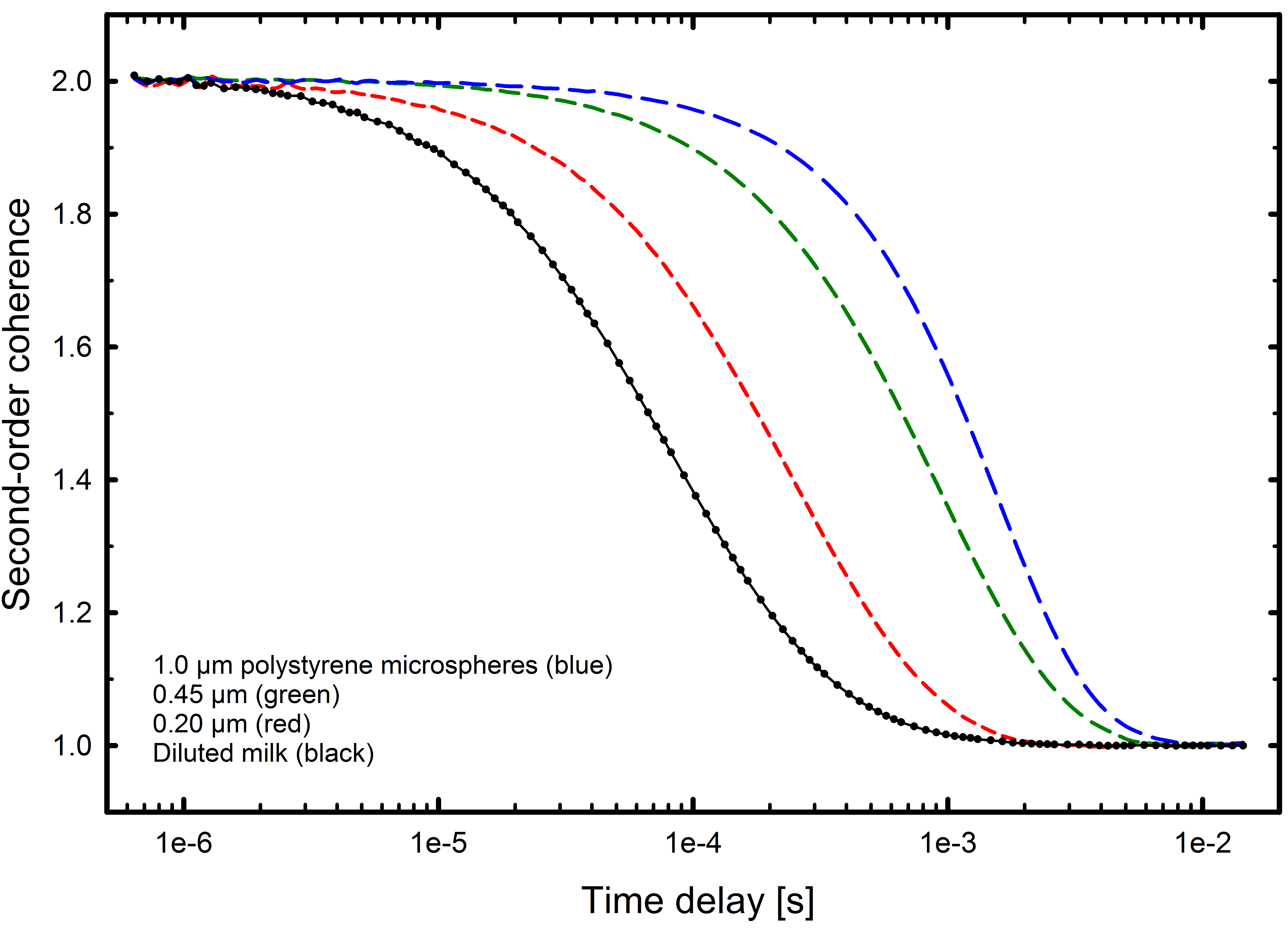}
      \caption{Measured second-order temporal coherence of light used to illuminate the artificial `stars'.  Laser light was scattered from suspensions of plastic microspheres undergoing thermal (Brownian) motion in water.  The coherence falls to unity for delays longer than the characteristic coherence times.  These are longer for larger particles (curves with longer dashes), which move more slowly, inducing less Doppler broadening.  Analogous functions are seen for diluted milk (solid curve with all measured data points), where microscopic fat globules act as scatterers.  These measurements of temporal coherence were made with a single telescope on a small and spatially coherent source.}
         \label{FigMicrospheres}
   \end{figure}
%

   \begin{figure*}
   \centering
   \includegraphics[width=\hsize]{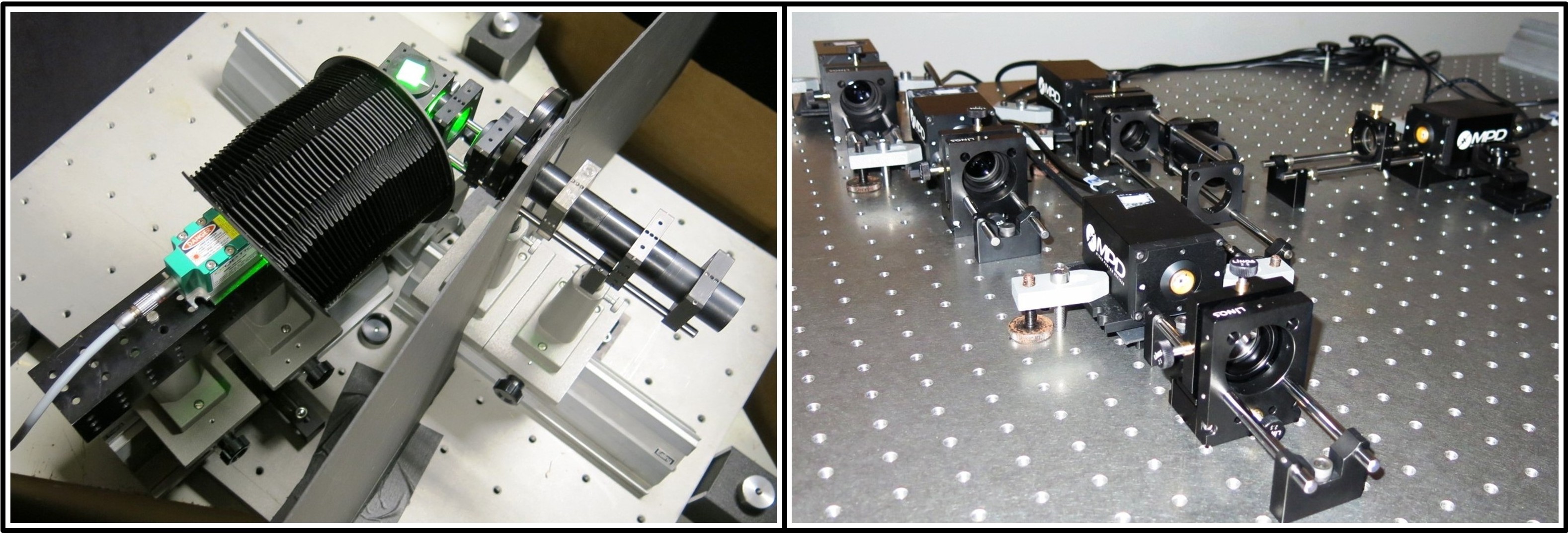}
      \caption{Components in the laboratory setup. Left: Light from a 300 mW $\lambda$~532 nm laser is made chaotic by scattering from microscopic particles in a square-top cuvette and focused by a condenser onto artificial `stars', which are mechanical apertures in a rotatable holder. Right: The `stars' are observed by a group of (here) five small telescopes with 25 mm apertures, each equipped with a photon-counting SPAD detector. One unit perpendicular to the others uses a 45-degree mirror to obtain a particularly short baseline. Another pair of two telescopes behind one beamsplitter serves to measure zero baseline, as required for calibrations.  Two-dimensional coverage is achieved by successively rotating the position angle of the source relative to the plane of the telescopes.}
     \label{FigOptics}
    \end{figure*}

\subsection{Dynamic light scattering}

Monochromatic light scattered by microscopic particles in thermal motion is broadened by Doppler shifts, producing a spectral line with a Lorentzian wavelength shape, and with a Gaussian (thermal, chaotic, maximum-entropy) distribution of the electric field amplitudes \citep{Berne_Pecora_2000, Crosignani_etal_1975}. This light is thus equivalent in its photon statistics and intensity fluctuations to the thermal (white) light expected from any normal star, the difference being that the spectral passband is now very much narrower. In addition to a high T$_{\rm{eff}}$, the narrowness of the spectral line implies a long coherence time, lessening the requirements on the detector time resolution and permitting measurements also on (the less demanding) microsecond scales, where detector concerns such as afterpulsing are less pronounced. The exact values for the spectral broadening can be modified by choosing the particle size (larger particles undergo slower motion, inducing less Doppler broadening), but it also somewhat depends on the viscosity and the temperature of the medium of suspension.

Such light sources are not completely without problems, however.  Intensity interferometry relies upon a relation between first- and second-order coherence that is valid for chaotic light. The scattering from single and non-interacting particles in Brownian motion produces this light,  but only in the case of single scattering. In the cases where the photons undergo multiple scattering events, if the scattering particles experience mutual interactions, where they have intrinsic dynamic properties, or where there are internal currents in the suspension liquid, this relation may no longer precisely hold. In such cases, there will still be relations between $\varg^{(1)}$ and $\varg^{(2)}$, but possibly in a somewhat different form. To avoid such complications, the optical paths were minimized by focusing the laser light very close to the surface of the fluid inside the cuvette.

Chaotic light was thus produced by scattering $\lambda$~532 nm laser light from microscopic monodispersive polystyrene spheres (from {\it{ Polysciences}}) suspended in a cm-sized cuvette with room-temperature distilled water.  After trying out spheres of several different sizes (ranging between 50 nm and 90 $\mu$m), a diameter of 0.2 $\mu$m was chosen as giving both an acceptably high light level in the scattering, and a convenient coherence time for measurements (Figure 2).  

However, the experiments required quite a number of tests, where the cost of volumes of such microscopic spheres started to become a problem. As an inexpensive scattering fluid with comparable properties, homogenized household milk (diluted with two parts water) was then used instead. Milk contains microscopic fat globules, which undergo analogous Brownian motion, also studied with photon correlation spectroscopy \citep{Alexander_Dalgleish_2006, Holt_etal_1973, Robin_Paquin_1991}. Given that the cost per volume is orders of magnitude lower than that of microsphere suspensions, several of the initial experiments were carried out with laser light scattering in milk.  Also, milk with varying fat content was tried out, but with no obvious differences in terms of, say, correlation timescales. Presumably, the homogenization process produces fat globules of similar size, and the difference between milk with differing fat content lies in its quantity, not its microscopic properties.  

The final optical setup was preceded by several experiments and optical ray-tracing simulations.  Only a relatively narrow range of parameters was found, where laboratory measurements appeared practical, in particular constrained by achievable source brilliance.  The laser was focused by optics into a very small volume in the scattering fluid, and that small illuminated volume was focused by a large condenser onto the tiny aperture of the artificial star, its brightness optimized with a three-axe micrometer.  For example, smaller artificial stars would have permitted a less compact and more easily handled interferometer array.  However, decreasing the pinhole diameter by just a factor 2, say, not only decreases the area and photon flux for the artificial star by a factor 4, but also diffracts its light over a twice greater angle, thus four times greater area, leading to an illumination decrease in the telescope plane by a factor 16.  Since the signal for intensity correlation is proportional to intensity squared, the observed signal would decrease by a factor 256, leading to impractical integration times.  

The chosen telescopes were a compromise between larger sizes to collect more photons and smaller ones to fully resolve the Fourier-plane structures.  Again, the functional dependences are steep: decreasing the telescope diameter a factor of 2, decreases its light-collecting area a factor 4 and the correlation signal by 16.  With careful optical design and using a more powerful laser, one should be able to increase the source brilliance, but probably not by a very large factor: focusing many watts into the submillimeter volume of the scattering fluid might induce currents, clog the microspheres, induce bubbles or even cause boiling. 

\subsection{Interferometer setup}

Figure 3 shows some photographs of the laboratory setup. The telescopes are small refractors with 25 mm diameter achromatic objective lenses mounted on an optical bench at some 23 meters distance from the artificial ‘stars’.  In each telescope, light is focused onto a SPAD, a single-photon-counting avalanche photodiode, enabling photon-count rates up to $\sim$10 MHz. The pulse-train output (electronic TTL standard) is fed to a computer-controlled firmware correlator for real-time cross computation of the data streams between various pairs of telescopes.  For more details, see \citet{Dravins_Lagadec_2014}. 

\subsection {Measurements of single `stars'}

Examples of measured intensity cross correlation functions between a pair of telescopes, $ \langle I_1(t) \cdot I_2(t+ \Delta t) \rangle $, are in Figure 4, normalized to $\varg^{(2)}$ = 2 for zero telescope baseline.  For long delay times $\Delta$t, the correlation vanishes, and the values tend to unity, as is appropriate for random uncorrelated variations. The characteristic slope for delays around $ \Delta t \sim$10$~\mu$s is a measure of the temporal coherence time and indicates the optical bandpass of the scattered laser light, here on the order of 10$^5$ Hz. For broader band light, the coherence time is much shorter, and then the possibilities of resolving the varying correlation for successively shorter delays become limited. However, in this arrangement it is possible to follow the detailed changes in both spatial and temporal correlations, when the temporal averaging proceeds over successively larger numbers of coherence times. 
 
For intensity interferometry we are not concerned about the temporal coherence but only the spatial one: its gradual change with increasing baseline enables the angular size of the source to be determined. At delays shorter than $\sim$1 $~\mu$s, the temporal coherence is here ‘fully’ resolved, and the differences between curves then reflect the spatial coherence only. For zero baseline, the spatial coherence must approach  $\varg^{(2)}$ = 2, and the autocorrelation function is normalized accordingly since several effects contribute to the actually measured values being lower.  For example, a slight deviation from zero baseline occurs because the telescope dimensions are not completely negligible and their measured spatial coherence corresponds to the convolution of the telescope pupils with the spatial coherence pattern \citep{Rou_etal_2013}. The ideal value of  $\varg^{(2)}$ = 2 is retrieved only for fully polarized light, while in the unpolarized light used here, the signal decreases by a factor $\sqrt{2}$ \citep{Hanbury_Brown_1974}.  Also, the measured signal is diluted by some (small) fraction of coherent laser light that did not scatter from the suspended microparticles but illuminated the aperture by being refracted in the walls of the glass cuvette, preserving its second-order coherence  $\varg^{(2)}$ = 1. To account for these effects, the successive cross correlation functions were normalized relative to the autocorrelation ones. Figure 5 shows the dependences of the measured second-order coherence on telescopic baseline.

   \begin{figure}
   \centering
   \includegraphics[width=\hsize]{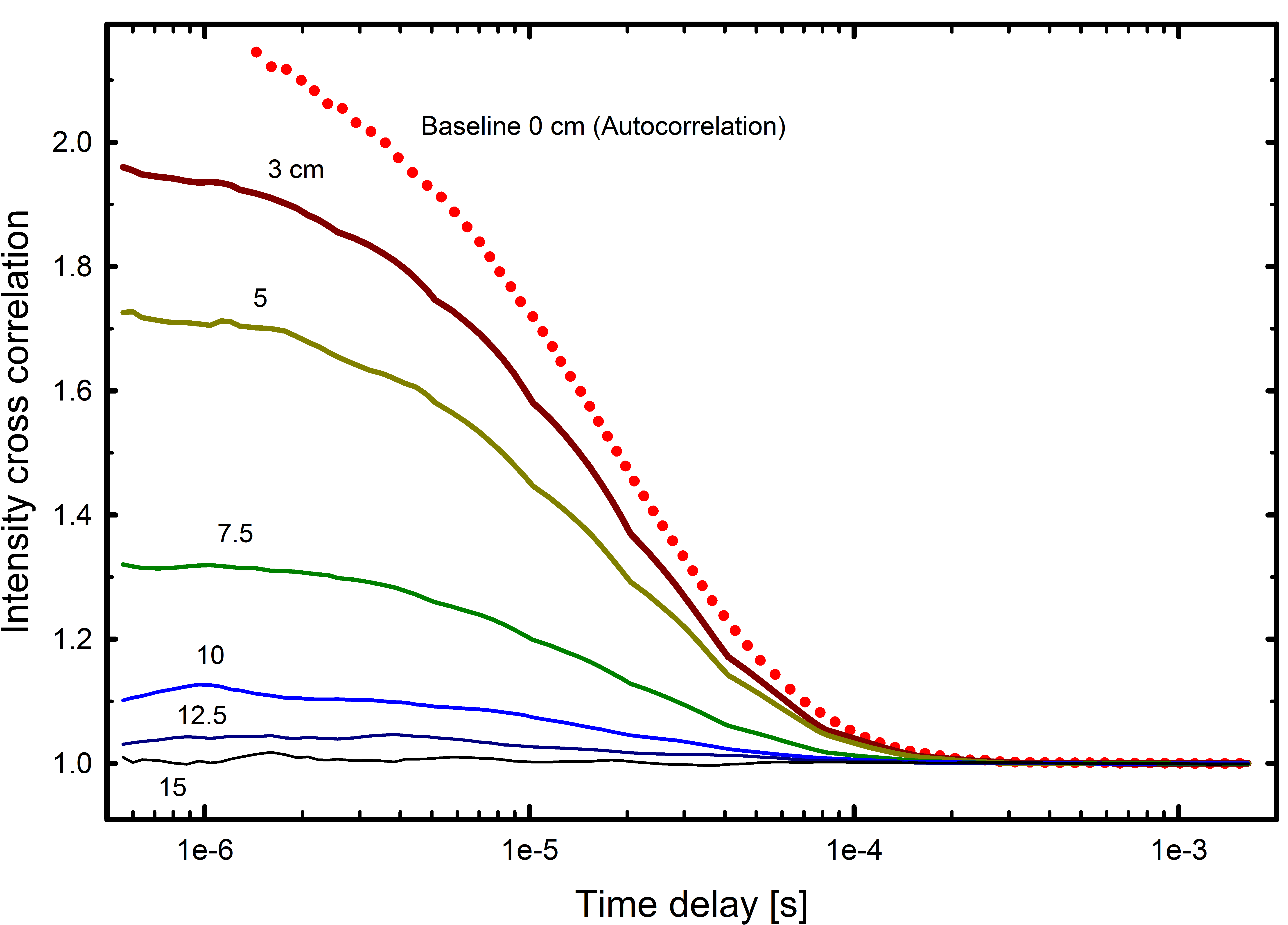}
      \caption{Cross correlation functions measured for an artificial single star of 1 arcsec apparent diameter (100 $\mu$m aperture at 23 m), normalized to zero baseline, show how (a) the temporal coherence gradually decreases with increasing time delay, and (b) the spatial coherence gradually decreases with increasing baseline.  In this particular case, the normalized value for zero baseline was taken as the autocorrelation for delays in the range 1-10 $\mu$s (at much shorter delay times the autocorrelation rises to unphysical values due to detector afterpulsing).  These effects were avoided in other measurements where the zero baseline was realized by cross correlating two telescopes behind one beamsplitter.}
         \label{FigCorrelations}
   \end{figure}
%

   \begin{figure}
   \centering
   \includegraphics[width=\hsize]{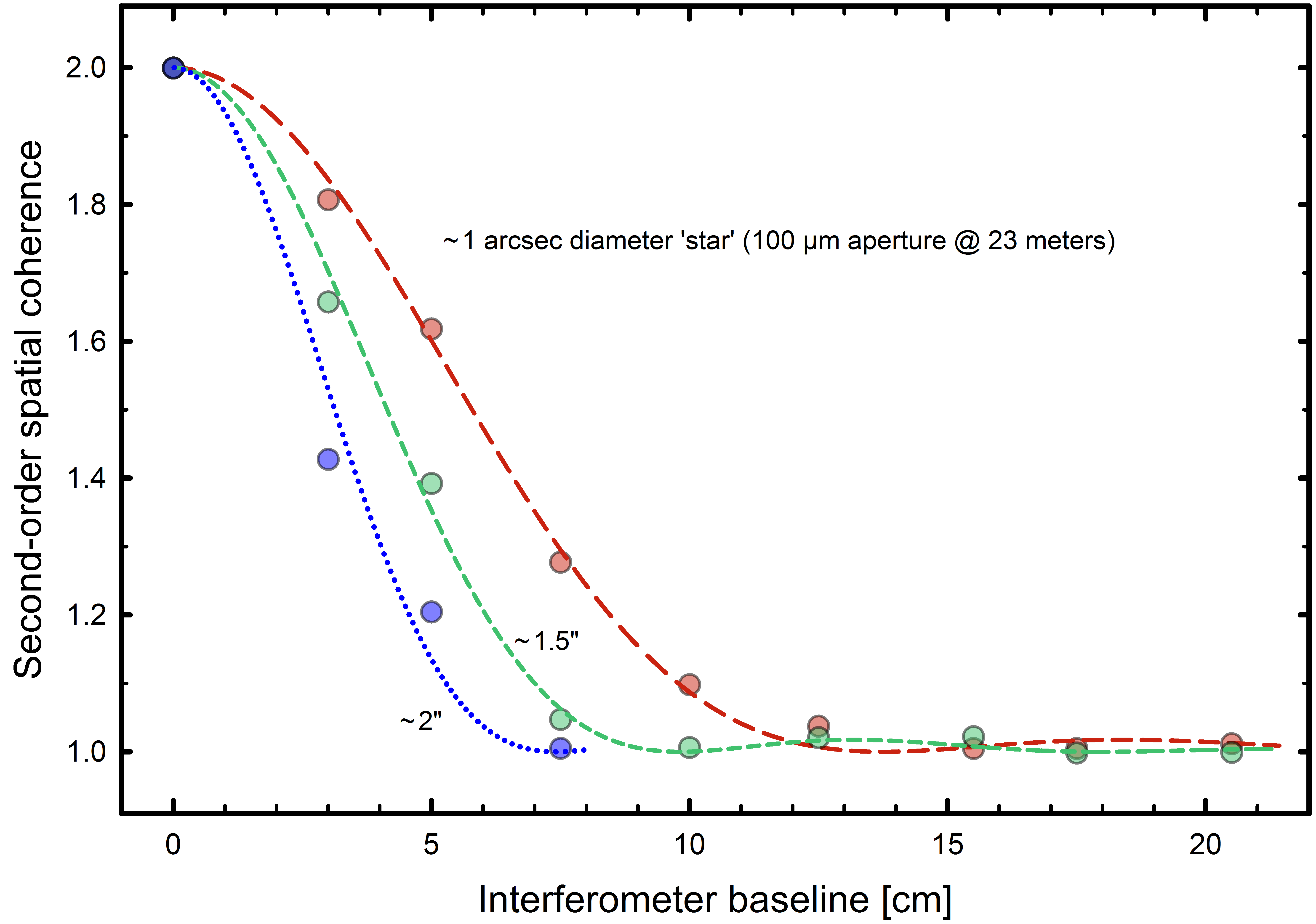}
      \caption{Second-order spatial coherence  $\varg^{(2)}$ measured for artificial single stars of different angular sizes. The coherence values come from correlation functions such as in Figure 4, averaged for delays between 1-10 $\mu$s, normalized to baseline zero. Superposed are theoretical dependences for ideal circular apertures (Airy patterns; squared moduli of the Fourier transforms, 1+ (2 $J_1(x)/x)^2$), confirming the expected trends.  The precise effectively illuminated source shapes depend on focusing details and are not exactly known; however typical measuring reproducibilities were within $\sim$0.02 of the coherence values.
}
         \label{FigCoherencePrinciple}
   \end{figure}
%

   \begin{figure}
   \centering
   \includegraphics[width=\hsize]{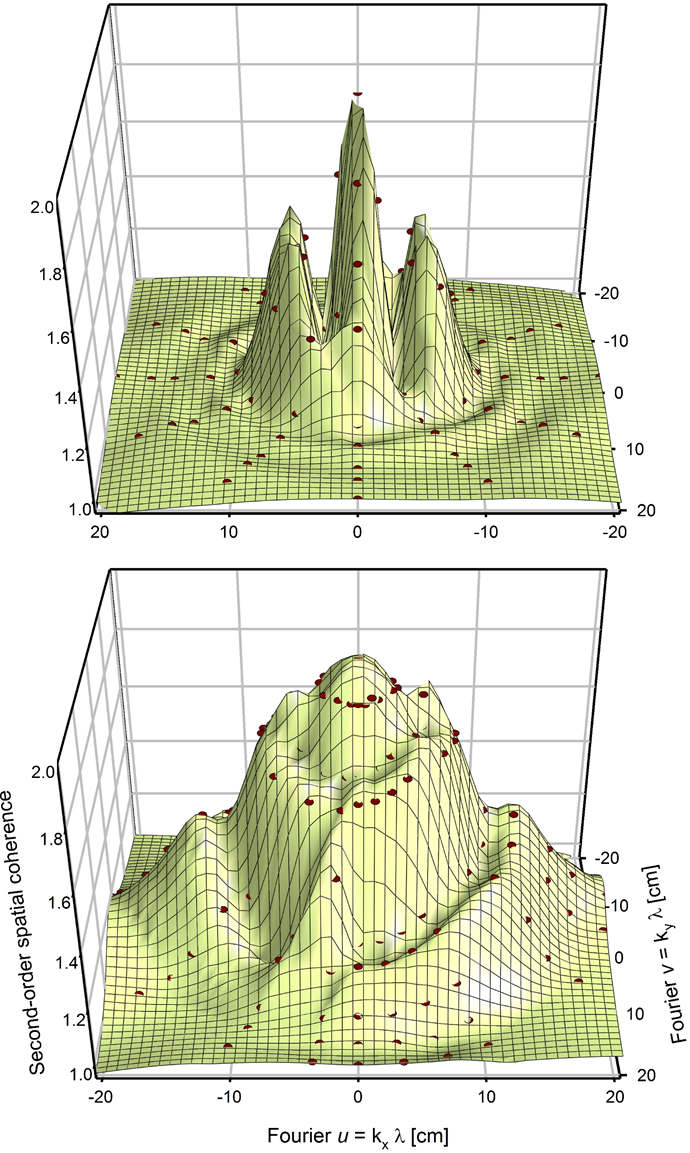}
      \caption{Second-order spatial-coherence surfaces $\varg^{(2)}$ measured at $\lambda$~532 nm for artificial stars in the laboratory.  Top: Binary `star' with equal components of diameter $\sim$1 arcsec, from intensity correlations measured between pairs of telescopes across 60 different non-redundant baselines.  The pattern of central maxima indicates the binary separation while the symmetric rings reveal the size of individual stars.  Black dots mark individual data points, also used in the later image reconstructions (the surfaces merely connect these measured points).  Bottom: A somewhat irregular and elliptical `star' of just below 1 arcsecond, measured over 100 baselines.  The data span the interferometric (u,v)-plane of the Fourier transform of the source image with different spatial (angular) wavenumbers {$\it{k}$}, corresponding to different telescope baseline lengths and orientations.  These maps provide the [moduli of the] Fourier transforms of the source images, analogous to the first-order diffraction patterns of intensity produced in coherent light (Figure 1).  Images reconstructed from these data are in Figure 8.}
         \label{3D_coherence}
   \end{figure}

\subsection{Hundreds of baselines: Mimicking large arrays}

A number of more complex sources measured included asymmetric stars and binary sources of varying sizes and separations. To mimic large telescope arrays, the effective number of telescopes and baselines was increased by making successive series of measurements across different angles in the interferometric (u,v)-plane. The telescope array is mounted horizontally and thus only covers horizontal baselines. To effectively cover oblique and vertical baselines as well, the artificial star was rotated to successively different position angles to change the (u,v)-plane coverage.  With N telescopes at non-redundant mutual separations, N(N--1)/2 different baselines can be constructed.  With five telescopes, ten baselines are available at each angular position and, for example, ten different position angles produce 100 different baselines.  The second-order coherence in a point (u,v) of the Fourier-transform plane equals the one in (--u,--v), so each baseline provides two points. Figure 6 shows coherence patterns for sources measured over 60 and 100 baselines, and Figure 7 such from 180 optical baselines. 

With such numbers, one moves into new parameter domains for optical interferometry, beginning to approach the capabilities of large Cherenkov arrays.  Assuming 50 telescopes will be available for interferometry, the number of baselines, N(N--1)/2, would already exceed 1000 (even if some might be redundant due to repetitive telescope locations).  As verified in numerical simulations, the ensuing nearly complete (u,v)-plane coverage enables full two-dimensional image reconstruction \citep{Nunez_etal_2012a, Nunez_etal_2012b}. One additional requirement during astronomical observing (although not needed here) will be to electronically and continuously track the changing projected baselines between pairs of telescopes as the source moves overhead across the sky during an observing night, assigning measurements to the instantaneous location in the (u,v)-plane. 

\section{Image reconstruction}

Image reconstruction from second-order coherence implies some challenges not present in first-order phase/amplitude techniques, and remains an active subject of research.  While intensity interferometry has the advantage of not being sensitive to phase errors in the optical light paths, it also does not measure such phases but instead obtains the absolute magnitudes of the respective Fourier-transform components of the source image.  These data can fit model parameters such as stellar diameters or binary separations but images cannot be computed through merely an inverse Fourier transform. 

Some strategies that specialize in analyzing intensity interferometry data have been developed; e.g.,  \citet{Holmes_Belen'kii_2004, Nunez_etal_2012a}, and consist in estimating Fourier phases from Fourier magnitudes.  Since the Fourier transform of a finite object is an analytic function, the Cauchy-Riemann equations can be used to find derivatives of the phase from derivatives of the magnitude, producing images that are unique except for translation and reflection.  Here, one could start with a one-dimensional phase estimate along a single slice through the (u,v)-plane origin.  Two-dimensional coverage requires combining multiple 1-D reconstructions, while ensuring mutual consistency between adjacent ones.  However, these algorithms require a very dense coverage of the (u,v)-plane and -- at least on first attempts -- did not prove efficient with current laboratory data.

Therefore, we chose to follow an inverse-problem approach, where a priori constraints are imposed to interpolate between successive Fourier frequencies. The imposed constraints are very general in character, and do not require much knowledge of the object. Such prior constraints (known as `regularization') are implemented in the `multi-aperture image reconstruction algorithm', MiRA: \citet{Thiebaut_2009}.  It has been tested for analyzing simulated intensity interferometry data \citep{Nunez_etal_2012b}, and we use it here.  The MiRA maximizes agreement with data, as well as with a general penalty (regularization) function. It can be stated as the following constrained optimization problem for an image $\mathbf{x}$:

\begin{equation}
  \mathbf{x}_{best}=arg\,min_x \left\{ f_{data}(\mathbf{x})+\mu f_{prior}(\mathbf{x})\right\}
\end{equation}

\noindent such that $x_k\geq 0$ for all $k$ and $\sum_{k=1}^N x_k=1$.\\

Here $\mathrm{x}$ is the image, $x_k$ the pixel value, and $\mu$ a constant. The function $f_{data}$ quantifies the agreement with the data as a  $\chi^2$, while $f_{prior}$ is the regularization function.  Here we use two different regularizations: smoothness and compactness, included in MiRA. The smoothness penalty function is defined as $f_{prior}=|\mathbf{x}-\mathbf{S} . \mathbf{x}|^2$, where $\textbf{S}$ is a smoothing operator implemented via finite differences. The compactness penalty is defined as $f_{prior}=\mathbf{x}^2$. The variable $\mu$ quantifies the global weight of the penalty function. The additional constraints are the positivity and normalizability of the image.  For reconstructing the image of the single elliptical star from the coherence patterns in Figure 6 (bottom), the compactness regularization was used with $\mu=10^{11}$, for the symmetric binary (Figure 6, top) the smoothness regularization was applied with $\mu=10^{14}$, and for the asymmetric one (Figure 7) $\mu=10^{13}$.  The resulting reconstructed images are in Figure 8, for which a shorter description appeared in \citet{Dravins_etal_2015}.  

Besides limits set by measurement noise, the fidelity of any reconstruction depends on how completely the (u,v)-plane has been filled with data.  In practice, every spatial frequency cannot be measured, and a fit to any sparse data set cannot yield a completely unique image without artifacts.  Of course, this also holds for any image reconstruction from a finite number of baselines in ordinary amplitude/phase interferometry.   

   \begin{figure}
   \centering
   \includegraphics[width=\hsize]{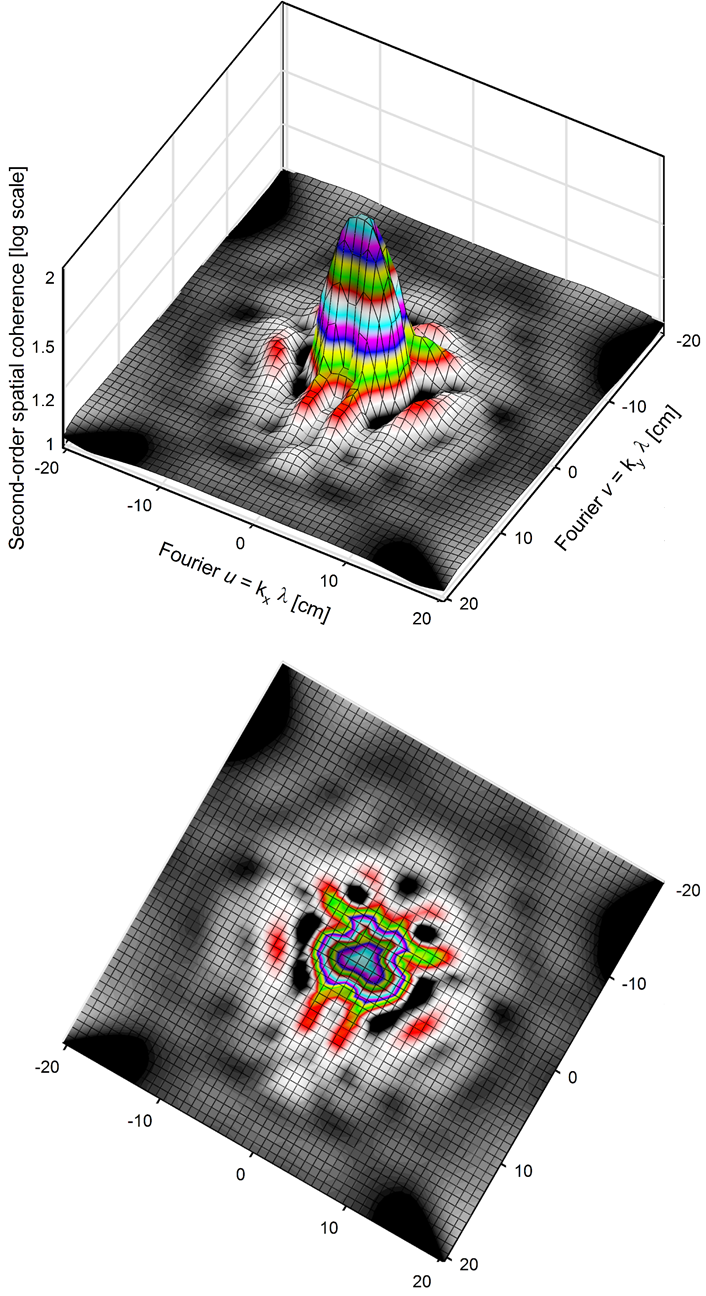}
      \caption{Second-order optical coherence of an artificial asymmetric binary star with differently sized components, built up from intensity-correlation measurements over 180 baselines between pairs of small optical telescopes.  The coordinates refer to the plane of the Fourier transform of the source image and correspond to different telescope baseline lengths and orientations.  At bottom, the projection of the 3-D mesh is oriented straight down, showing [the modulus of] the source’s Fourier transform (‘diffraction pattern’).  The image reconstructed from these data is in Figure 8, bottom.}
         \label{Fig_Asymmetric_coherence}
   \end{figure}
%

   \begin{figure}
   \centering
   \includegraphics[width=\hsize]{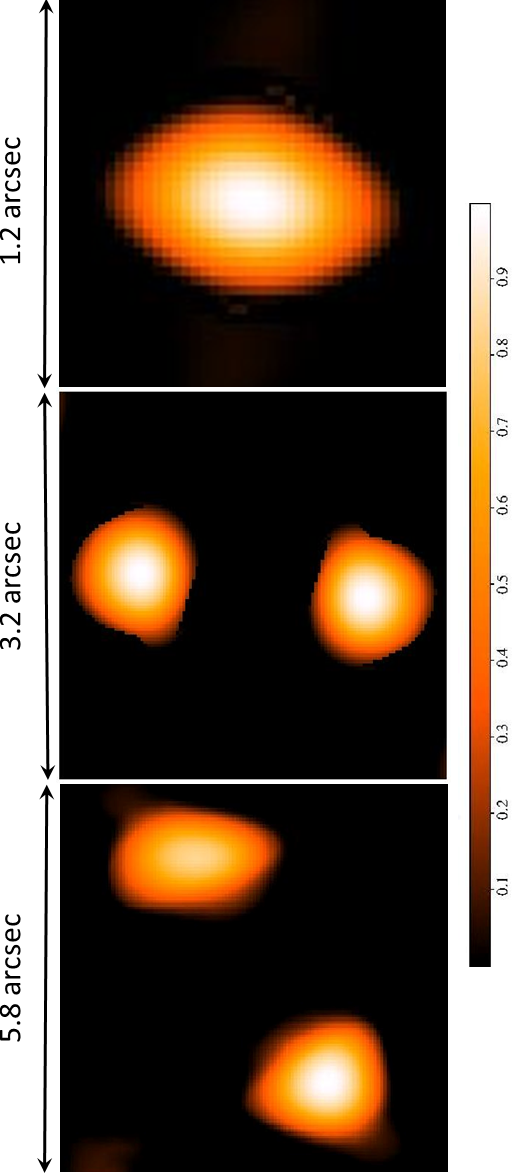}
      \caption{Images reconstructed from intensity interferometry measurements of (top to bottom) a single, somewhat irregular elliptical `star', a binary `star' with equal, and one with unequal components, from measurements with 100, 60, and 180 optical baselines, respectively.  Sufficiently dense coverage of the interferometric Fourier-transform plane enables a reconstruction of the full source images, despite the lack of Fourier phase information.  The scale on the right denotes relative intensities.  In principle, the images are unique except for their possible mirrored reflections.   These are believed to be the first diffraction-limited images that have been reconstructed from an array of optical telescopes, connected only through electronic software, with no optical links between them.}
         \label{Fig_Reconstruction}
   \end{figure}

\subsection{Image reconstruction practicalities}

The numerical stability and reproducibility of the reconstructions were examined.  For the single asymmetric star in Figure 8 (top), its semi-major axis was measured as 0.53 ${\pm}$~0.01, and its semi-minor one as 0.36 ${\pm}$~0.01 arcsec.  The `edge' of the star was here taken as pixels above 0.5\% of the image maximum, and the reproducibility estimated by initiating the image reconstruction with different random images.  However, the current setup does not permit a very precise comparison of the retrieved image to the original since the effective source shape is set by the illumination pattern falling onto the geometric aperture.  While the latter was independently measured in homogeneous laser illumination (Figure 1), the critical focusing of its illuminating scattered light probably produces some inhomogeneities across it.

In the case of the binary stars, one limitation comes from the finite (u,v)-plane coverage: with baselines between 3 and 20.5 cm, the former constrains the largest field to nominally 4.5 arcsec, and the latter limits the resolution to 0.65 arcsec.  Precise measures that can still be obtained include the angular separation of the symmetric binary, measured between the highest-valued pixels as 1.90 ${\pm}$~0.01 arcsec.  For the asymmetric binary, its flux ratio ($\approx$~4) is retrieved reasonbly well, but the relative sizes of the components are not.  This behavior is consistent with earlier simulations, which found that the precision is affected by the brightness ratio between both binary members, becoming an issue when one of them is more than about three times brighter than the other \citep{Nunez_etal_2012a}.  
 
The computing effort is modest.  Starting from a random image, and measurements across 100 baselines, a 10,000-pixel image is retrieved in some 10 seconds on a 3 GHz processor.  This time can be shorter if there already is some initial guess that somewhat resembles the final image (perhaps obtained from some `raw’ reconstruction from a Cauchy-Riemann analysis), or can be longer for additional baselines or more numerous pixels.  In any case, it is by far not comparable to the extensive numerical efforts involved in radio aperture synthesis that also involve spectral analysis for each spatial point.

The quality of the reconstructed images may depend on the density of numerical sampling across the Fourier plane.  For the reconstructions in Figure 8, no smoothing or interpolation of the measured data points were carried out.  Tests with various sampling densities indicated that for best image reconstruction, the number of (u,v)-plane samples should roughly equal the number of baseline measurements.  Possibly, other dependences could hold for noisier data, where some smoothing could be called for.  In general, however, the optimization of reconstruction in the presence of various noise sources and incomplete Fourier-plane coverage is a fairly complex problem that requires further study (analogous to what in the past has been done for aperture synthesis in radio).
 
Similar to any aperture synthesis from phase/amplitude interferometry, the fidelity of the reconstructed images depends on the density of measurements and on their extent across the (u,v)-plane, as well as the performance of any `cleaning’ algorithms applied to remove signatures of the (u,v)-plane sampling.  While such algorithms of course cannot create information content on spatial scales that are not measured, properly deconvoluted images will not display confusing instrumental signatures. 

\subsection{Triple correlations and `closure intensities'}

Besides the cross correlation required to retrieve the spatial coherence $\varg^{(2)}$, there is a potential to also explore higher-order correlations in light.  Current measures are obtained as cross correlations between intensity fluctuations at two spatial locations, at one instant in time. However, using telescope arrays, one can construct, for example, third-order intensity correlations, $\varg^{(3)}$ for systems of three telescopes: $\langle I(r_1, t_1) \cdot I(r_2, t_2) \cdot I(r_3, t_3) \rangle$, where the temporal coordinates  $t$ do not even have to be equal.  In principle, such higher-order correlations in light carry additional information about the source \citep{Fontana_1983, Gamo_1963, Malvimat_etal_2014, Marathay_etal_1994, Ofir_Ribak_2006, Sato_etal_1978, Sato_etal_1981}. For instance, from correlations among all possible triplets and quadruplets of telescopes, a more robust full reconstruction of the source image should be possible. Here, we are entering largely unexplored territory for observational astronomy, but one that could well be enabled by the availability of extended arrays with large and numerous telescopes. 

A two-point correlation measures the squared modulus of the complex degree of mutual coherence $\gamma_{ij}\gamma_{ij}^*$ between two telescope locations $i$ and $j$.  Analogously, it can be shown that three-point intensity correlations allow a quantity to be measured that is proportional to the product of the complex degree of mutual coherence between detectors in a closed loop, i.e., $\gamma_{ij}\gamma_{jk}\gamma_{ki}$ \citep{Malvimat_etal_2014}. By expressing $\gamma_{ij}$ as $\gamma_{ij}=|\gamma_{ij}|exp\{i(\phi_{ij}+\Delta_{ij})\}$, where $\phi_{ij}$ is the astrophysical Fourier phase, and $\Delta_{ij}$ is the atmospherically induced phase, it is straightforward to show that the atmospheric phases cancel out in the triple product. The phase of the triple product is a linear combination of the Fourier phases, and is known as the `closure-phase' in amplitude interferometry. Measuring the closure phase for many triplets in an interferometric array allows the Fourier phase across the array to be constrained, which may in turn enhance image reconstruction.  \citet{Malvimat_etal_2014} and \citet{Wentz_Saha_2015} investigated the S/N properties of three-point intensity correlations, showing that this quantity may be measurable for bright stars, while Nu\~nez \& Domiciano de Souza (in preparation) are investigating the use of three-point correlations for enhancing image reconstruction.  While three-point correlations may turn out to provide useful information about the Fourier transform of the source, these have more severe S/N limitations than do two-point ones. Therefore, their main role in imaging should be to supplement two-point data.  However, if such closure-phase intensity interferometry turns out to be useful, it could provide additional constraints at very little additional effort since the intensity signals from each telescope can be freely copied or even stored off-line for later analysis.

\section{Toward kilometer-scale optical interferometry}

The current laboratory setup of a multitelescope array has demonstrated the end-to-end operation of an imaging intensity interferometer of the Hanbury Brown-Twiss type, involving very many baselines and densely populating large areas of the interferometric Fourier (u,v)-plane. The photon count rates achieved and handled by the correlators in real time (>1 MHz) are comparable to what can be expected in actual stellar observations, with some spectral feature selected through a narrow bandpass filter.  Already our current correlators can handle signals from up to 20 telescopes simultaneously and, together with experiments carried out with actual Cherenkov telescopes \citep{Dravins_LeBohec_2008, LeBohec_etal_2010}, these experiences bode well for future full-scale observations of celestial sources.

\subsection{Pending laboratory issues}

The current experiments were carried out with thermal light, which is produced by laser scattering. Although the photon statistics are believed to closely match those of thermal white light, the coherence times are much longer than the subnanosecond scales relevant for actual astronomical sources. Since coherence times for broader-band light are very much shorter than realistic electronic timescales, most of the cross correlation signal will then be confined to the shortest temporal delays of timescales on a few nanoseconds. Here, various detector issues begin to appear, such as deadtimes and afterpulsing, whose understanding, calibration and mitigation may be essential \citep{Dravins_Lagadec_2014}. With the aim of realizing laboratory experiments also on such short timescales, some tests with high-temperature white-light lamps have already been made, indicating that it might be feasible to achieve meaningful S/Ns, although some effort will be required.

For the observability of fainter sources, the S/N can be improved by simultaneous measurements in multiple spectral bands, exploiting the property that S/N is independent of optical bandpass. Possibly, some scheme for dispersive optics could be devised, probably near the focal plane, followed by a multielement detector array; e.g., \citet{Gerwe_etal_2013}.   A better solution, however, would be to obtain detectors that are both energy-resolving and photon-counting.  While such devices have been demonstrated in the laboratory, they do not yet seem to be practical for field work with large telescopes.  Obviously, the highest quantum efficiency is desirable, such as values >90\% that have been demonstrated with superconducting nanowire devices for single-photon detection, reaching into the infrared.  Also polarization properties could be optimized: Since the full $\varg^{(2)}$ signal appears in polarized light, that could be exploited by introducing a polarizing beamsplitter and detecting each polarization separately. 

How many of these features that can be practically combined will determine what limiting magnitudes ultimately can be reached, with a potential for going substantially fainter than the m${_V}$=8 estimate for single-wavelength observations discussed in Section 3 above.  Further information is contained in various higher-order correlations, but these have not yet been experimentally studied in the context of astronomical imaging, and theoretical simulations are lacking here also.  

\subsection{Outlook for field experiments}

Cherenkov telescopes normally produce cm-size stellar images in their focal planes, which are much larger than the single-pixel SPAD detectors used here. Although our detectors are similar in principle to those large-area solid-state photomultipliers now demonstrated in Cherenkov telescopes \citep{Anderhub_etal_2013, Catalano_etal_2013, Sottile_etal_2013}, those larger area ones are built up by a large number of individual light-sensitive areas on the same silicon chip and (as we have examined in other laboratory experiments) possess somewhat different deadtime, dark-count, and afterpulsing characteristics. An adequate understanding of these detector properties (or of photomultipliers, in case such would be used) will be required to reach the photometric precisions required for disentangling physical values of $\varg^{(2)}$ from stochastic or systematic noise.

When observing celestial sources during an observing night with stationary telescopes, projected baselines between pairs of telescopes will gradually change. Of course, this enables an even richer (u,v)-plane coverage, and indeed it is the principle of aperture synthesis using Earth rotation \citep{Saha_2011}. However, to retain a high Fourier-plane resolution in real-time correlation, this requires an electronic unit that implements a variable time delay onto the stream of photon pulses, compensating for the relative timings of the wavefront at the different telescopes, as the source moves across the sky \citep{Dravins_etal_2012}. In case the correlation is computed off-line, following a recording of the signal during observation, the variable delay would instead have to be applied in software.

   \begin{figure}
   \centering
   \includegraphics[width=\hsize]{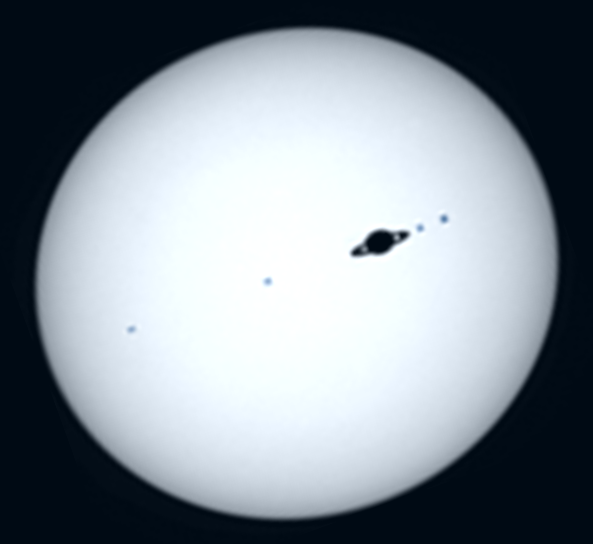}
      \caption{Vision of microarcsecond optical imaging: Expected resolution for an assumed transit of a hypothetical exoplanet across the disk of Sirius.  Stellar diameter = 1.7 solar; distance = 2.6 pc; angular diameter = 6 mas; assumed planet of Jupiter size and oblateness; equatorial diameter = 350 $\mu$as; Saturn-type rings; four Earth-size moons.  With a Cherenkov telescope array spanning 2 km, already a 50 $\mu$as resolution provides more than 100 pixels across the stellar diameter.}
         \label{FigExoplanet}
   \end{figure}

\section{Microarcsecond astrophysics}

\subsection{Observing with small Cherenkov telescope arrays}

Probably the `easiest' targets for intensity interferometry observations are relatively bright and hot, single or binary stars of spectral types O and B or Wolf-Rayet stars with their various circumstellar emission-line structures. Their stellar disk diameters are typically $\sim$0.2–0.5 mas, so they lie (somewhat) beyond what can be resolved with existing instruments.  For amplitude interferometry, limitations also arise because of the paucity of unresolved (but sufficiently bright) reference stars for calibrating fringe visibilities.  Intensity interferometry does not require such reference sources but merely instrumental stability between measurements made with shorter and longer baselines, very likely becoming critical in enabling baselines on the order of 1 km or more. 

Other candidate targets include rapidly rotating stars, with oblate shapes deformed by rotation, circumstellar disks, winds from hot stars, blue supergiants and extreme objects, such as $\eta$~Carinae, interacting binaries, the hotter parts of [super]nova explosions, pulsating Cepheids, or other hotter variables \citep{Dravins_etal_2012, Dravins_etal_2013, Trippe_etal_2014}.

Given that the S/N is independent of optical bandwidth, one might just as well observe in the light of some specific spectral feature, as in white light. Assuming suitable wavelength filters, one could try to map the non-radial pulsations across the surfaces of stars such as Cepheids. The amplitudes in temperature and in white light probably are modest but the associated velocity fluctuations might be observable. If the telescope optics can be collimated enough to permit the use of very narrow-band spectral filters centered on strong absorption lines, the local stellar surface will appear at their particular residual intensity when at rest relative to the observer, but will reach continuum intensity when local velocities have Doppler-shifted the absorption line outside the filter passband. Once such spatially resolved observations of stellar non-radial oscillations have been realized, they should provide significant input to models of stellar atmospheres and interiors.

\subsection{Targets for large Cherenkov telescope arrays}

The full Cherenkov Telescope Array is foreseen to have its numerous telescopes distributed over a few square kilometers, with an edge-to-edge distance of two or three km \citep{Acharya_etal_2013, CTA_2015}. If fully equipped for intensity interferometry at the shortest optical wavelengths, the spatial resolution will approach $\sim$30 $\mu$as. Such resolutions have hitherto been reached only in radio, and it is problematical to speculate on what features could appear in the optical.  However, to appreciate the meaning of these resolutions, Figure 9 shows an `understandable' type of object: a hypothetical exoplanet in transit across the star Sirius (T$_{\rm{eff}}$ = 9,940 K).  The planet size and oblateness was taken as equal to that of Jupiter, but fitted with a Saturn-like ring and four ‘Galilean’ moons, here assumed to be Earth-sized.  Such an exoplanet would subtend a readily resolvable angle of some 350~$\mu$as.  While spatially resolving the disk of an exoplanet in its reflected light may remain unrealistic for the time being, the imaging of its dark silhouette on a stellar disk – while certainly still very challenging – might not be impossible \citep{Strekalov_etal_2013, Strekalov_etal_2014}.

Intensity interferometry actually possesses some advantages for these possible observations. The lack of sensitivity to the phases of the Fourier components of the image could be an advantage here since one would always measure `only' the amplitude of the Fourier transform of the exoplanet image, irrespective of where on the stellar disk it happens to be. Unless close to the stellar limb, the star only serves as a bright background, its comparatively `huge' diameter of 6 mas not contributing any sensible spatial power at any relevant telescopic baselines.

\begin{acknowledgements}

This work was supported by the Swedish Research Council and The Royal Physiographic Society in Lund. The development of concepts for intensity interferometry with Cherenkov telescope arrays has involved interactions with several colleagues elsewhere, in particular at the University of Utah in Salt Lake City (David Kieda, Stephan LeBohec, and others) and at the University of Padova (Cesare Barbieri, Giampiero Naletto, and others). Early experiments toward laboratory intensity interferometry at Lund Observatory involved also Toktam Calv{\'e}n Aghajani, Hannes Jensen, Ricky Nilsson and Helena Uthas while laboratory studies of SPAD detectors were also made by Daniel Faria and Johan Ingjald. The artificial `stars' were prepared by the late research engineer Nels Hansson.  We thank Colin Carlile for careful proofreading of of the manuscript.  This paper was completed while DD was a visiting scientist at the European Southern Observatory in Santiago de Chile.   

\end{acknowledgements}

\end{document}